\documentclass[%
reprint,amsmath,amssymb,aps,
superscriptaddress,
preprintnumbers,
nofootinbib,
 amsmath,amssymb,
 aps,
]{revtex4-2}

\usepackage{graphicx}
\usepackage{dcolumn}
\usepackage{bm}
\usepackage{hyperref}
\usepackage{cleveref}
\usepackage{booktabs}

\usepackage{physics}
\usepackage[english]{babel}
\usepackage{color}
\usepackage{xr}
\usepackage{subfigure}

\newcommand{\ee}[1]{\times 10^{#1}}

\begin{document}

\title{Direct Detection of Hawking Radiation from Asteroid-Mass Primordial Black Holes}

\author{Adam Coogan}
\email{a.m.coogan@uva.nl }
\affiliation{GRAPPA, Institute of Physics, University of Amsterdam, 1098 XH Amsterdam, The Netherlands}

\author{Logan Morrison}%
\email{loanmorr@ucsc.edu}
 \affiliation{Department of Physics, University of California, Santa Cruz, CA 95064, USA}

\author{Stefano Profumo}%
\email{profumo@ucsc.edu}
\affiliation{Department of Physics, University of California, Santa Cruz, CA 95064, USA}

\date{\today}

\begin{abstract}
Light, asteroid-mass primordial black holes, with lifetimes in the range between hundreds to several millions times the age of the universe, are well-motivated candidates for the cosmological dark matter. Using archival COMPTEL data, we improve over current constraints on the allowed parameter space of primordial black holes as dark matter by studying their evaporation to soft gamma rays in nearby astrophysical structures. We point out that a new generation of proposed MeV gamma-ray telescopes will offer the unique opportunity to directly detect Hawking evaporation from observations of nearby dark matter dense regions and to constrain, or discover, the primordial black hole dark matter.
\end{abstract}

\maketitle
\clearpage

Discerning the fundamental nature of the cosmological dark matter (DM) is perhaps the most pressing issue in particle physics. While much is known about the average density of DM in the universe as a whole as well as on the density of DM in specific structures, the mass of its elementary constituent is largely unconstrained. Roughly, macroscopic quantum effects, such as the DM featuring a De Broglie wavelength comparable to the size of the smallest gravitationally collapsed structures, bound the DM mass from below. The stability of structures such as galactic disks and the existence of relatively old, stable, unperturbed systems such as globular clusters and dwarf spheroidal galaxies constrains the DM mass from above, since a large point mass would measurably disrupt such structures. Most of the roughly ninety orders of magnitude in between these two model-independent constraints could well accommodate the mass of the elementary objects making up (most of the) DM (see~\cite{Profumo:2017hqp} for a review).

Essentially all information about what the DM is therefore stems from gravitational interactions. As far as observations are concerned, the DM need not be ``charged'' under any other additional interaction besides gravity. An extensive experimental and observational program has for many years assumed that the DM is charged under the Standard Model's weak nuclear interactions. This program, however, has thus far failed to bear positive fruit. The class of DM candidates known as weakly interacting massive particles (WIMPs), while remaining solidly theoretically motivated, does not appear to be supported by any uncontroversial experimental evidence~\cite{Arcadi:2017kky}.

The successful detection of gravitational waves~\cite{Abbott:2016blz} has ushered a renewed interest in DM candidates with masses of the order of the black holes whose binary mergers were directly detected~\cite{Bird:2016dcv}. The somewhat surprising mass and spin distributions of such black holes, as inferred from observations, brought further momentum to the notion that the black holes might in fact not be all of stellar origin, but that some (or all) of them be of {\em primordial} origin~\cite{Fernandez:2019kyb} (see e.g.~\cite{Carr:2020gox,Green:2020jor} for recent reviews on primordial black holes, or PBHs). At the same time, closer scrutiny of the range of viable masses for PBHs has unveiled that previously-thought excluded regions are in fact perfectly viable~\cite{Green:2020jor}.

Stellar-mass black holes, such as those whose binary mergers are detected via gravitational wave telescopes, could well be a significant fraction of the DM. Constraints from CMB distortions~\cite{Poulin_2017, Clark_2017} and dynamical effects on small-scale structures~\cite{Brandt:2016aco} are subject to significant debate and systematic issues, while constraints dependent on their merger rates might also have been overestimated (see e.g. the recent study~\cite{Boehm:2020jwd}). Lighter black holes with horizon sizes comparable to visible light and masses around $10^{-11}\ M_\odot$ or $10^{22}$ grams are constrained by microlensing of stars. Again, recent work has shown how finite-size source effects must be very carefully taken into account to avoid overestimating the constrained parameter space~\cite{Smyth:2019whb}.

Much lighter black holes are extremely challenging to detect. Femtolensing constraints~\cite{Barnacka_2012}, employing much shorter wavelengths than visible light, turned out to also have neglected the impact of finite source size~\cite{Katz:2018zrn,Montero_Camacho_2019} and do not set any meaningful constraints. Destruction of white dwarfs and neutron stars was also found to be plagued by issues with the black hole capture rate, and does not set any strong constraints at present (see e.g.~\cite{Montero_Camacho_2019}).

Lighter and lighter black holes have increasingly large Hawking temperatures ($T_H\approx (10^{10} \ {\rm g}/M)\ {\rm TeV}$) and evaporate much more efficiently and quickly, with a lifetime $\tau \approx 10^{66}(M/M_\odot)^3$ years. Black holes lighter than $\approx5\times 10^{14}$ g have a lifetime comparable to the age of the universe, while slightly more massive black holes are currently evaporating. Constraints can thus be set from searches for the evaporation products of these $10^{16}-10^{17}~\mathrm{g}$ holes, assuming they are a fraction $f_{\rm PBH}$ of the cosmological DM. Evaporation of black holes at all redshifts and in all structures can be constrained by the requirement not to overproduce the extragalactic gamma-ray background (EGRB)~\cite{Carr:2009jm}. Evaporation can also lead to CMB distortions~\cite{Poulin_2017,Clark_2017}, heating of neutral hydrogen~\cite{Clark:2018ghm}, and of the interstellar medium in dwarf galaxies~\cite{Kim:2020ngi}. The local density of PBH, in the mass range where evaporation is significant, is also constrained by measurements of the abundance of electrons and positrons in the cosmic radiation from Voyager 1~\cite{Boudaud:2018hqb}. Positrons from evaporation are additionally constrained by the 511 keV annihilation line with electrons as observed by INTEGRAL~\cite{DeRocco:2019fjq,Laha:2019ssq}. INTEGRAL data also directly constrain the abundance of PBH in the Galaxy, as shown in Ref.~\cite{Laha_2020}. Finally, there exist constraints from the diffuse neutrino background as measured by Super-Kamiokande from evaporation to neutrinos~\cite{Dasgupta:2019cae}.

In this work, we find that observations with COMPTEL give the {\em strongest constraints currently available} over a broad range of black hole masses. We study the prospects for discovering these PBHs with next-generation MeV gamma-ray telescope observations of the Milky Way, Andromeda (M31) and nearby dwarf spheroidal galaxies. In deriving these constraints we present a robust, semianalytical calculation of the secondary photon spectrum from evaporating PBHs with MeV-scale temperatures. This is required for correctly assessing the sensitivity of telescopes to PBHs at the low end of the mass range we consider.

The remainder of this study is as follows: after describing the current observational status, we list future telescopes relevant for the detection of Hawking radiation, and describe the salient features that would enable detection of black hole evaporation. We then describe the details of Hawking evaporation and its detection, present our results, and conclude.

\begin{figure}
    \centering
    \includegraphics[width=0.5\textwidth]{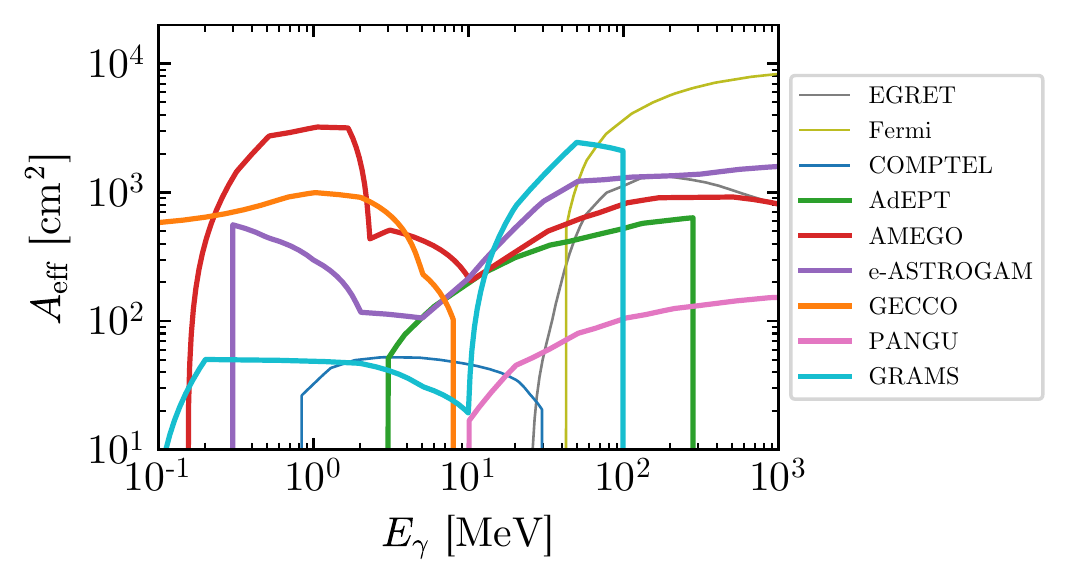}
    \caption{\textbf{The effective area, as a function of energy, of existing and proposed MeV gamma-ray telescopes.} Thin lines mark existing telescopes and thick lines mark proposed ones. The effective area of MAST (not shown) ranges from $\sim 7\ee{4} - 10^5\, \mathrm{cm}^2$ over $E_\gamma = 10\, \mathrm{MeV} - 60\, \mathrm{GeV}$.}
    \label{fig:a_eff_spec}
\end{figure}

\vspace*{.5cm}

\noindent\emph{MeV gamma-ray telescopes.} The Hawking temperature $T_H$ of interest for black holes of mass $M$ whose lifetimes are within a few orders of magnitude of the age of the universe $\tau_U$ falls in the MeV scale:
\begin{equation}
    \tau(M)\simeq 200\ \tau_U \left( \frac{M}{10^{15}\ {\rm g}} \right)^3 \simeq 200\ \tau_U \left( \frac{10\ {\rm MeV}}{T_H} \right)^3.
\end{equation}
PBH with a Hawking temperature in the GeV would have a lifetime of less than $3\times10^6$ years. At present they cannot comprise a significant fraction of the cosmological DM since that would imply too large a DM abundance at early times, in conflict with CMB and BBN observations. Instead, PBHs evaporating at present are generically expected to be producing photons in the MeV range. This limits the available observational capabilities relevant for constraining PBH evaporation to the low-energy range of the Fermi Large Area Telescope (Fermi-LAT)~\cite{fermi}, and to its predecessors EGRET~\cite{egret} and COMPTEL~\cite{Kappadath:1993} on board the Compton Gamma Ray Observatory, and the INTErnational Gamma-Ray Astrophysics Laboratory (INTEGRAL)~\cite{Winkler:2003nn}. We show in \cref{fig:a_eff_spec} the relevant effective areas, in cm$^2$, as a function of energy, with solid lines.

Several missions with capabilities in the MeV are in the proposal, planning, or construction phase. Here, we consider the following: AdEPT~\cite{adept}, AMEGO~\cite{amego}, e\-ASTROGAM~\cite{e_astrogam}\footnote{This has since been scaled back to All-Sky-ASTROGAM~\cite{as_astrogam}.}, GECCO~\cite{gecco_loi,alex_slides}, MAST~\cite{mast}, PANGU~\cite{pangu,pangu_aeff} and GRAMS~\cite{grams,grams_loi}. For future missions, we will assume dedicated observation times of $T_\mathrm{obs} = 10^8\, \mathrm{s} \approx 3\, \mathrm{yr}$. We note that in searching for Hawking evaporation products, energy and angular resolution are not critical. The spectra, to be discussed below, consist of a fairly broad peak with a long, low-energy tail. As long as the target's angular size is larger than the telescope's angular resolution, the latter does not enhance detection capabilities either.

\begin{figure*}
    \centering
    \includegraphics[width=0.9\textwidth]{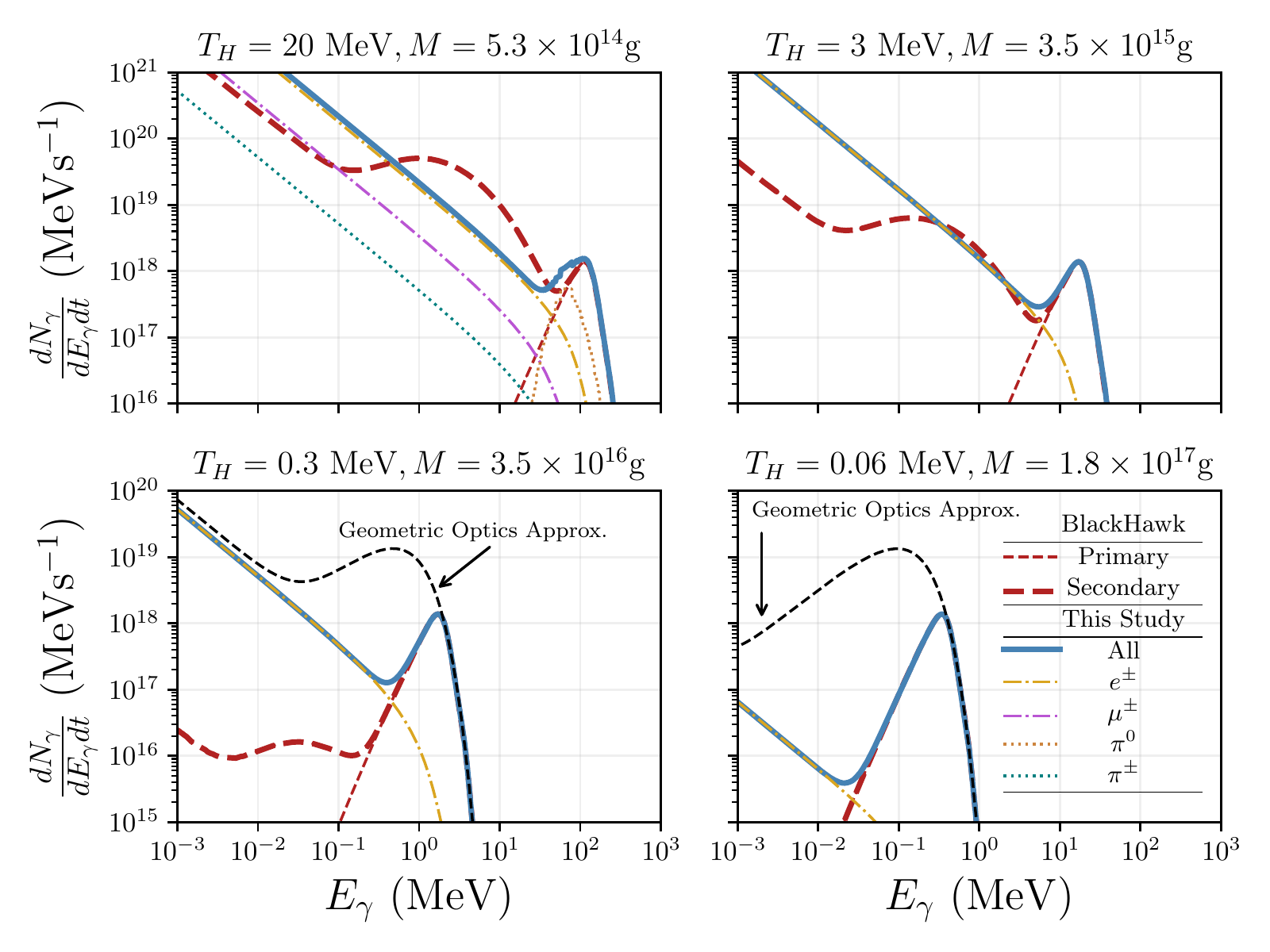}
    \caption{\textbf{Photon emission from light black hole evaporation.} We consider Hawking temperatures of 20, 3, 0.3 and 0.06 MeV (from top left to bottom right), corresponding to masses $M = 5.3\times 10^{14},\ 3.5\times 10^{15},\ 3.5\times 10^{16},\ 1.8\times 10^{17}$ grams. The thick blue lines show the spectra computed in this work; the dashed red curves correspond to the primary (thin lines) and secondary-plus-primary (thick lines) output from {\tt BlackHawk}. We also show contributions from $\pi^0$ decay (magenta dotted lines) and from final state radiation off of electrons and muons (dot dashed yellow and magenta lines) and charged pions (dotted green lines). In the two lower panel we also show what results from adopting the geometric optics approximation for the grey-body factors.}
    \label{fig:spectra}
\end{figure*}

\vspace*{.5cm}

\noindent\emph{Photons from evaporating PBHs.} A non-rotating black hole with mass $M$ and corresponding Hawking temperature $T_H=1/(4\pi G_N M)\simeq 1.06 (10^{16}\ {\rm g}/M)\ {\rm MeV}$, with $G_N$ Newton's gravitational constant, emits a differential flux of particles per unit time and energy given by (in natural units $\hbar=c=1$)
\begin{equation}\label{eqn:hawkrad}
    \pdv{N_i}{E_i}{t} = \frac{1}{2\pi} \frac{\Gamma_i(E_i, M)}{e^{E_i/T_H}-(-1)^{2s}},
\end{equation}
where $\Gamma_i$ is the species-dependent grey-body factor, and $E_i$ indicates the energy of the emitted particle of species $i$. Unstable particles decay and produce stable {\em secondary} particles, including photons. The resulting differential photon flux per solid angle from a region parametrized by angular direction $\psi$ is obtained by summing the photon yield $N_\gamma$ from all particle species the hole evaporates to:
\begin{equation}
    \dv{\phi_\gamma}{E_\gamma} = \frac{1}{4\pi} \int_{\mathrm{LOS}} dl\ \pdv{N_\gamma}{E}{t} \frac{f_{\rm PBH}\, \rho_{\rm DM}(l, \psi)}{M}.
\end{equation}
We assume the PBHs have a monochromatic mass function, comprise a fraction $f_{\rm PBH}$ of the DM and trace the DM's spatial distribution.

Notice that upon integrating over the appropriate solid angle, the expression above contains a factor identical to what found in decaying DM searches, which we denote by
\begin{equation}
    \bar{J}_D \equiv \frac{1}{\Delta\Omega} \int_{\Delta\Omega} d\Omega \int_{\mathrm{LOS}} dl\ \rho_{\rm DM}(r(l, \psi)).
\end{equation}
We list $\bar{J}_D$ in \cref{tab:J_factors} for the inner $5^\circ$ of the Milky Way, Draco, and M31, assuming the PBH spatial density is described by a Navarro-Frenk-White (NFW) profile~\cite{NFW}. We find that a $5^\circ$ observing angle provides a close-to-optimal balance of signal to background. To bracket uncertainties in the Galactic DM distribution, we also consider the possibility that it follows an Einasto profile~\cite{einasto}. We include the $\bar{J}_D$ factor from galactic PBHs for the region $|b| < 20^\circ,\, |\ell| < 60^\circ$ observed by COMPTEL~\cite{Kappadath:1993,Essig:2013goa}, assuming an NFW Galactic DM halo.

\renewcommand{\arraystretch}{1.2}
\begin{table}[]
    \centering
    \begin{tabular}{cc}
        \toprule
        Target & $\bar{J}_D\, (\mathrm{MeV}\, \mathrm{cm}^{-2}\, \mathrm{sr}^{-1})$  \\
        \midrule
        Draco (NFW)~\cite{Dugger:2010ys} & $1.986\ee{24}$\\
        M31 (NFW)~\cite{Geehan:2005aq} & $4.017\ee{24}$\\
        Galactic Center (NFW)~\cite{deSalas:2019pee}  & $1.597\ee{26}$\\
        Galactic Center (Einasto)~\cite{deSalas:2019pee} & $2.058\ee{26}$\\
        $|b| < 20^\circ,\, |\ell| < 60^\circ$ (NFW)~\cite{deSalas:2019pee} & $4.866\ee{25}$\\
        \bottomrule
    \end{tabular}
    \caption{\textbf{$\bar{J}_D$ factors for various circular targets and the COMPTEL observing region from Ref.~\cite{Kappadath:1993}.} The DM profile parameters are taken from the indicated references. For the Milky Way targets, we use the values from Table III of Ref.~\cite{deSalas:2019pee}. The Einasto profile parameters are adjusted within their $1\sigma$ uncertainty bands to maximize $\bar{J}_D$. For all other targets we use the parameters' central values. The distance from Earth to the Galactic Center is set to 8.12 kpc~\cite{Abuter:2018drb,deSalas:2019pee}. For reference, the angular extent of a $5^\circ$ region is $2.39\ee{-2}\, \mathrm{sr}$.}
    \label{tab:J_factors}
\end{table}

To generate the gamma-ray spectra from a decaying PBH, we employ the greybody factors calculated by the publicly available code {\tt BlackHawk}~\cite{arbey2019BlackHawk}. {\tt BlackHawk} generates primary spectra for all fundamental SM particles using the standard Hawking evaporation spectrum given in \cref{eqn:hawkrad}. The code also uses {\tt PYTHIA} \cite{Sjostrand:2006za} and {\tt HERWIG} \cite{Bahr:2008pv} \footnote{{\tt BlackHawk} has options of use either {\tt PYTHIA} or {\tt Herwig} at the BBN epoch (using concrete lifetimes for various unstable particles) or {\tt PYTHIA} at the present epoch, where all unstable particles are decayed. We exclusively use {\tt PYTHIA} at the current epoch.} to hadronize and shower strongly-interacting and unstable particles, producing the full (primary and secondary) spectra of all stable SM particles. However, the hadronization routines in both of these codes are only reliable for energies $\gtrsim 5$ GeV. In fact, {\tt BlackHawk} uses {\em extrapolation tables} to compute spectra from particles with energies below the range where {\tt PYTHIA} and {\tt HERWIG} are computed for, which result in unreliable and unphysical spectra.\footnote{The {\tt BlackHawk} authors are aware of this, and the code raises the following warning when a user attempts to compute the secondary spectra below 5 GeV: {\tt WARNING ENERGY BOUNDARIES ARE NOT COMPATIBLE WITH THE NEW PYTHIA HADRONIZATION TABLES! NEW PYTHIA HADRONIZATION TABLES WERE COMPUTED FOR 5.00e+00 GeV < E < 1.00e+05 GeV EXTRAPOLATION WILL BE USED}.}

In addition, since {\tt PYTHIA} is designed for collider physics, it rejects photons which are sufficiently collinear to the radiating charged particle. This is because events in which the photon and charged particle are not well separated cannot be distinguished from events with no photon in collider detectors. However, on cosmic scales, the propagation lengths of the photon and charged particle are large enough to completely separate the two, making {\tt PYTHIA}'s isolation cut too restrictive.

\begin{figure}
    \centering
    \includegraphics[width=0.45\textwidth]{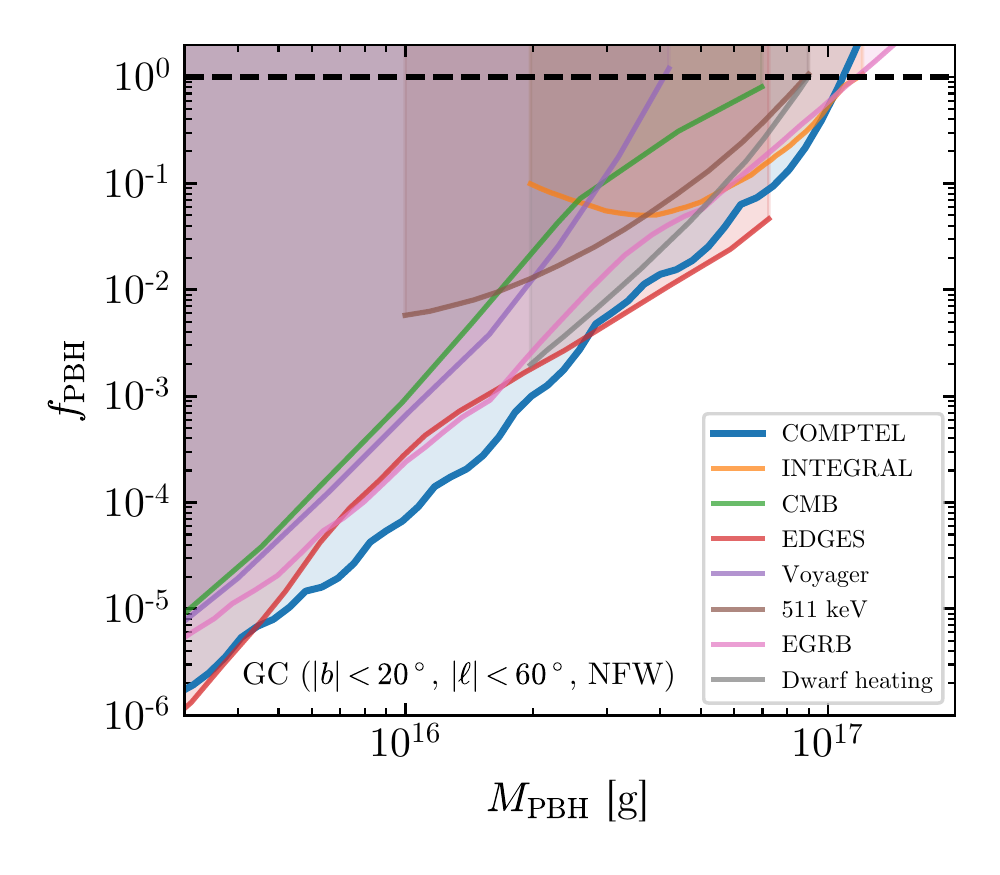}
    \caption{\textbf{New bounds on the PBH abundance based on COMPTEL observations of the Milky Way.} Assuming the PBHs follow an Einasto rather than NFW distribution gives a slightly stronger bound. Existing bounds collected in Ref.~\cite{Green:2020jor} from INTEGRAL observations of galactic diffuse emission~\cite{Laha_2020}, CMB~\cite{Poulin_2017,Clark_2017}, EDGES 21 cm~\cite{Clark:2018ghm}, Voyager 1~\cite{Boudaud:2018hqb}, the 511 keV gamma-ray line~\cite{DeRocco:2019fjq,Laha:2019ssq}, the extragalatic gamma-ray background~\cite{Carr:2009jm} and dwarf galaxy heating~\cite{Kim:2020ngi} are also shown.}
    \label{fig:bounds_comptel}
\end{figure}

\begin{figure*}
    \centering
    \includegraphics[width=0.9\textwidth]{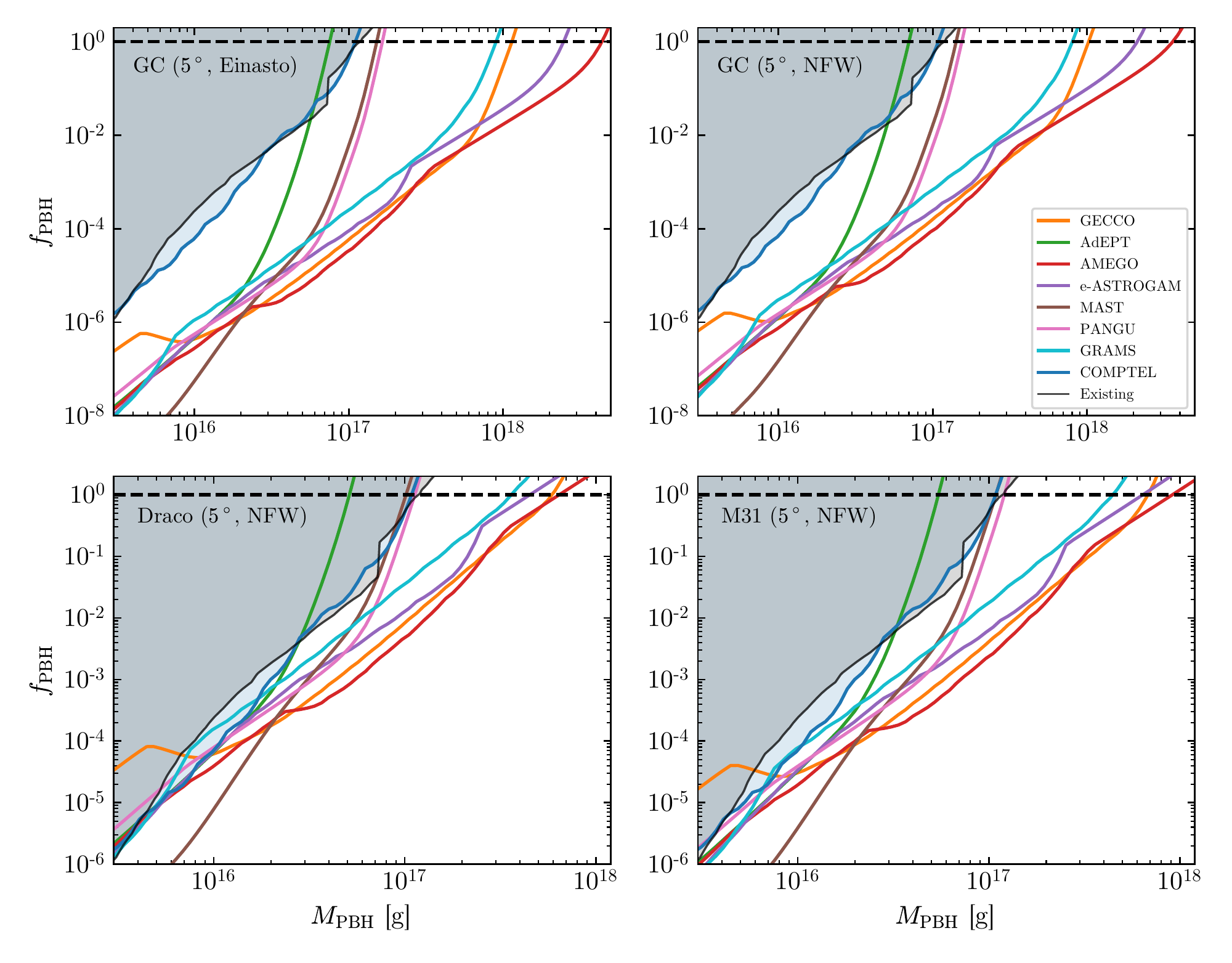}
    \caption{\textbf{Discovery reach for PBHs with future MeV gamma ray telescopes.} In the top row, the PBH density is assumed to track an Einasto density profile (left) and an NFW profile (right) fit to the Milky Way's DM distribution. In the lower panels we consider the Draco dwarf spheroidal and M31 as targets. For all targets we assume a $5^\circ$ circle around the central region.}
    \label{fig:discovery_reach}
\end{figure*}

Instead, we use {\tt BlackHawk} to generate primary spectra of photons, electrons and muons, and we use {\tt BlackHawk}'s tables of greybody factors to compute the primary Hawking radiation of neutral and charged pions. We model the final-state radiation off the charged final state particles by ``convolving'' the primary spectrum with the Altarelli-Parisi splitting functions at leading order in the electromagnetic fine-structure constant $\alpha_{\mathrm{EM}}$\cite{Chen:2016wkt,Altarelli:1977zs}. For the unstable particles, such as pions, we use {\tt Hazma}~\cite{hazma} to compute the photon spectrum from decays. Our total photon spectrum is then:
\begin{align}
    &\pdv{N_{\gamma}}{E_{\gamma}}{t} = \pdv{N_{\gamma,\mathrm{primary}}}{E_{\gamma}}{t}\\ 
    &\quad + 
    \sum_{i=e^{\pm},\mu^{\pm},\pi^{\pm}}\int\dd{E_{i}}
    \pdv{N_{i,\mathrm{primary}}}{E_{i}}{t}
    \dv{N^{\mathrm{FSR}}_{i}}{E_{\gamma}}\notag\\
    &\quad + \sum_{i=\mu^{\pm},\pi^{0},\pi^{\pm}}\int\dd{E_{i}}
    \pdv{N_{i,\mathrm{primary}}}{E_{i}}{t}
    \dv{N^{\mathrm{decay}}_{i}}{E_{\gamma}},\notag
\end{align}
where the FSR spectra are given by:
\begin{align}
    \dv{N^{\mathrm{FSR}}_{i}}{E_{\gamma}}&= \dfrac{\alpha_{\mathrm{EM}}}{\pi Q_f}P_{i\to i\gamma}(x)\qty[\log(\dfrac{(1-x)}{\mu_{i}^2})-1]\notag,\\
    P_{i\to\gamma i}(x)& = \begin{cases}
        \frac{2(1-x)}{x}, & i=\pi^{\pm}\\
        \frac{1+(1-x)^2}{x}, & i=\mu^{\pm},e^{\pm}\\
    \end{cases},
\end{align}
with $x=2E_{\gamma}/Q_f$, $\mu_{i}=m_{i}/Q_f$ and $Q_f=2E_f$. (See~Ref.~\cite{hazma} for explicit expressions of $dN^{\mathrm{decay}}/dE_{\gamma}$ for the muon, neutral and charged pions.) 

We illustrate the issues mentioned above in \cref{fig:spectra}, where we show secondary spectra computed with {\tt BlackHawk}, which, as mentioned, include unphysical extrapolations of the QCD fragmentation results outside their range of validity, evident in the unphysical bumps at low energy. Note that the bump is likely a remnant of what is expected from neutral pion decay. However, on a log scale in energy the emission from $\pi\to\gamma\gamma$ is symmetric around $m_{\pi^0/2}$, which is not the case in the extrapolated spectra. Additionally, we note that while the bump over- or undershoots the actual photon emission, the asymptotic low-energy behavior is also incorrect, as explained above, because of the lack of soft collinear photons. Finally, we note that for the largest masses/lowest Hawking temperatures, the treatment of final state radiation off of electrons and positrons leads to a significant underestimate of the emission in the keV range. We also show in the two lower panels the spectra that would result from adopting the geometric optics approximation, as done e.g. in Ref.~\cite{Boudaud:2018hqb}, which leads to a significant overestimate of the particles' fluxes. In this current work, correctly accounting for the secondary spectrum impact constraints on low-mass PBHs from telescopes sensitive to low-energy gamma rays (e.g. GECCO; c.f. \cref{fig:a_eff_spec}).

\vspace*{.5cm}

\noindent\emph{COMPTEL Bounds \& discovery reach.} To set constraints with COMPTEL data, we find the largest value of $f_\mathrm{PBH}$ such that the photon flux from PBHs in the region $|b| < 20^\circ,\, |\ell| < 60^\circ$ does not exceed the observed flux plus twice the upper error bar in any energy bin:
\begin{equation}
    \left[ \int_{E_\mathrm{low}^{(i)}}^{E_\mathrm{high}^{(i)}} dE_\gamma\ \dv{\Phi_\gamma}{E_\gamma} \right] \leq \Phi_\gamma^{(i)} + 2 \Delta\Phi_\gamma^{(i)}, \, i = 1, \dots, n_\mathrm{bins}.
\end{equation}
The integral ranges from the lower to upper bound of each bin, indexed by $i$. This procedure yields conservative limits since it makes no assumptions about the astrophysical background. However, with background modeling we expect the constraints to improve by less than an order of magnitude~\cite{Essig:2013goa}.

For analyzing the discovery potential for future telescopes, we require the signal-to-noise ratio over the observing period to be larger than five: $N_\gamma|_\mathrm{PBH} = 5 \sqrt{N_\gamma|_\mathrm{bg}}$. Given a signal or background flux $\frac{d\Phi}{dE}$, the number of photons detected is given by
\begin{equation}
    N_\gamma = T_\mathrm{obs} \int_{E_\mathrm{min}}^{E_\mathrm{max}} dE_\gamma\ A_\mathrm{eff}(E_\gamma) \, \int dE_\gamma'\ R_\epsilon(E_\gamma | E_\gamma') \dv{\Phi}{E_\gamma}.
\end{equation}
Here $A_\mathrm{eff}$ is the energy-dependent telescope's effective area (c.f. \cref{fig:a_eff_spec}). The function $R_\epsilon(E | E')$ is a Gaussian with mean $E'$ and standard deviation $\epsilon(E')\, E'$ that accounts for the telescope's finite energy resolution. We ignore energy dependence in $T_\mathrm{obs}$.

For targets oriented away from the Galactic center (Draco and M31), we adopt an empirical power law background model fit to high-latitude COMPTEL and EGRET data~\cite{Boddy:2015efa}:
\begin{equation}
    \dv{\Phi_\gamma}{E_\gamma} = 2.74\ee{-3} \left( \frac{E}{\mathrm{MeV}} \right)^{-2}\, \mathrm{cm}^{-2}\,\mathrm{s}^{-1}\,\mathrm{sr}^{-1}\,\mathrm{MeV}^{-1}. \label{eq:default_bg}
\end{equation}
In the case of targets focused on the Galactic center, we use a more sophisticated model~\cite{Bartels:2017dpb}. It consists of bremsstrahlung, $\pi^0$ and inverse-Compton spectral components computed with {\tt GALPROP}~\cite{Strong:2009xj} and calibrated against data in the window $|\ell| < 30^\circ\, |b| < 10^\circ$, as well an additional power law component required to fit COMPTEL data. This flux predicted by this model is roughly a factor of $7$ larger than in \cref{eq:default_bg}.

We carry out this analysis by implementing a new model for PBH dark matter in our code {\tt hazma}~\cite{hazma}.

\vspace*{.5cm}

\noindent\emph{Results and discussion} The PBH abundance bound we derive from COMPTEL data is displayed in \cref{fig:bounds_comptel}, along with a host of existing evaporation constraints. The COMPTEL bound is the most stringent constraint by a factor of $\sim 3$ for PBH masses near $10^{16}\, \mathrm{g}$ and in line with other constraints over the rest of the mass range we consider.

The discovery reach for selected planned MeV gamma-ray telescope using observations of the Galactic center, Draco and M31 are shown in \cref{fig:discovery_reach}. We highlight that AMEGO, e-ASTROGAM and GECCO observations of the Galactic center are capable of discovering PBH DM up to a mass of $\sim 10^{18}\, \mathrm{g}$, an order of magnitude larger than current constraints. Note that neglecting the secondary evaporation spectrum computed above would lead to underprojecting GECCO's discovery reach by an order of magnitude at the lower bound of the mass range in our plots. All of the experiments considered herein could discover PBHs with an abundance an order of magnitude below current constraints in part of the mass range $5\ee{15} - 3\ee{18}\, \mathrm{g}$. We emphasize that having a low energy threshold is important for pushing the discovery reach into the asteroid mass window, as can be seen by comparing the effective areas in \cref{fig:a_eff_spec} with the curves in \cref{fig:discovery_reach}. Due to the relative large observing region ($5^\circ$), these projections are not particularly sensitive to whether the Galactic PBH distribution follows an Einasto or NFW profile. In the case of M31 or Draco observations we predict a fainter signal, but expect PBHs with masses up to $\sim 10^{18}\, \mathrm{g}$ to be discoverable. 

\vspace*{.5cm}

\noindent\emph{Summary \& conclusions.} We considered bounds on the fractional contribution that primordial black holes with lifetimes comparable to the age of the universe make to the cosmological dark matter. We pointed out that since the relevant Hawking temperature is around the MeV scale, computing their secondary evaporation spectra requires appropriately treating the final state radiation off of charged leptons and light hadrons, as well the production and decays of light mesons. We showed that at present and across a large swath of black hole masses the best constraints stem from COMPTEL observations of the central region of the Galaxy. We considered an optimistic range of possible future telescopes with MeV-band coverage, and pointed out that many of those will have a distinct opportunity to discover Hawking evaporation from evaporating PBH making up a large fraction of the DM with masses in the $10^{17}\lesssim M / \mathrm{g} \lesssim {\rm few}\times 10^{18}$. Direct detection of black hole evaporation would have enormous consequences for the quest to discern the nature of the cosmological DM, for understanding the early universe, and for black hole physics.

\vspace*{.5cm}

\begin{acknowledgments}
LM and SP are partly supported by the U.S.\ Department of Energy grant number de-sc0010107.
\end{acknowledgments}

\bibliographystyle{unsrt}

\bibliography{references}

\end{document}